\documentclass[12pt]{article}

\usepackage{amsmath}
\usepackage{amssymb}
\usepackage{epsfig}

\newcommand{\set}[1]{{\mathbb{#1}}}
\newcommand{\one}{\mbox{\tt 1}\hspace{-0.057 in}\mbox{\tt l}}
\newcommand{\Tr}{\mbox{\rm Tr}}

\begin{document}

\title{Bayesian updating of a probability distribution encoded
  on a quantum register}

\author{
Andrei N. Soklakov and R\"udiger Schack\\
\\
{\it Department of Mathematics, Royal Holloway,
       University of London,}\\
{\it Egham, Surrey TW20 0EX, United Kingdom}}

\date{15 November 2005}
\maketitle

\begin{abstract}
  We investigate the problem of Bayesian updating of a probability
  distribution encoded in the quantum state of $n$ qubits.  The updating
  procedure takes the form of a quantum algorithm that prepares the quantum
  register in the state representing the posterior distribution. Depending on
  how the prior distribution is given, we describe two implementations, one
  probabilistic and one deterministic, of such an algorithm in the
  standard model of a quantum computer.
\end{abstract}

\section{Introduction}

Bayes's rule provides a simple and fundamental mechanism for updating a
probability distribution in the light of new data~\cite{Bernardo1994}. 
The rule takes its simplest
form for a finite sample space, $\set{H}$, where the elements $h\in\set{H}$
can be identified with the atomic events, or {\em hypotheses}. Let $P_{\rm
  prior}(h)=P(h)$ be the prior probability distribution, and assume some piece
of data, $d$, is observed. If $P(d|h)$ is the conditional probability of $d$,
given $h$, Bayesian updating consists of replacing the prior with the
posterior distribution, $P_{\rm posterior}=P(h|d)$, where
\begin{equation} \label{ConditionalPD}
P(h|d)=\frac{P(d|h)P(h)}{\sum_{h}P(d|h)P(h)}\;.
\end{equation}

To simplify the notation, we assume from now on that the set of hypotheses is
of the form $\set{H}=\{0,\dots,2^n-1\}$ for some positive integer $n$. 
For $h\in\set{H}$, let $|h\rangle$ denote the computational basis states of
a register of $n$ qubits. The state
\begin{equation}  \label{PriorState}
|\Psi_{\rm prior}\rangle
=\sum_{h\in\set{H}}\sqrt{P(h)}\,|h\rangle
\end{equation}
provides an encoding of the prior on the quantum register. Even though the size
of the sample space grows exponentially with the number of qubits, $n$, there
exists an interesting class of priors for which $|\Psi_{\rm prior}\rangle$ can
be prepared efficiently, in the sense that the required computational resources
grow only polynomially with $n$ \cite{Grover-0208,Soklakov2005b}. 

To formulate the problem of Bayesian updating for a prior encoded on a quantum
register, we make the assumption that we have a classical algorithm that
computes, as a function of $h$, the conditional probability $P(d|h)$ for the
observed data $d$.  Given this classical algorithm, the goal of Bayesian
updating is then to prepare the register in the state
\begin{equation} \label{PosteriorState}
|\Psi_{\rm posterior}\rangle=\sum_{h\in\set{H}}\sqrt{P(h|d)}\,|h\rangle\;,
\end{equation}
with $P(h|d)$ given by Eq.~(\ref{ConditionalPD}). 
If the prior is given to us in the form of a single copy of the state  
$|\Psi_{\rm prior}\rangle$, our problem is equivalent to finding a quantum
operation, $M_d$, that maps any prior 
state of the form~(\ref{PriorState}) into the 
corresponding posterior state of the form~(\ref{PosteriorState}),
\begin{equation}
M_d|\Psi_{\rm prior}\rangle=|\Psi_{\rm posterior}\rangle\;.
\end{equation}
It is easy to see that $M_d$ cannot in general be a trace-preserving map.
For example, consider the two prior states 
\begin{equation} \label{ExamplePriors}
|\Psi_{\rm prior}^1\rangle=\frac{1}{\sqrt{2}}(|1\rangle+|2\rangle)\,,\ \ \ 
|\Psi_{\rm prior}^2\rangle=\frac{1}{\sqrt{2}}(|2\rangle+|3\rangle)\,,
\end{equation}
corresponding to two different prior probability distributions, 
and assume that the conditional probability distribution is given by
\begin{equation}
P(d|h)=\left\{\begin{array}{ll}
               0 & {\rm if\ }h= 2\,,\cr
               c\neq 0 & {\rm otherwise}\,,
                  \end{array}
                   \right.
\end{equation}
where $c$ is a constant determined by normalization.
Although the prior states (\ref{ExamplePriors}) 
are nonorthogonal, we obtain mutually orthogonal 
posterior states
\begin{equation}
|\Psi_{\rm posterior}^1\rangle=M_d|\Psi_{\rm prior}^1\rangle=|1\rangle\;,\;\;\;
|\Psi_{\rm posterior}^2\rangle=M_d|\Psi_{\rm prior}^2\rangle=|3\rangle\;,
\end{equation}
which implies that $M_d$ is trace-decreasing. Bayesian updating of a single
copy of $|\Psi_{\rm prior}\rangle$ is therefore generally probabilistic.
Section II of this paper discusses probabilistic Bayesian updating. 

A deterministic updating scheme is possible, however, if the prior is given in
the form of a unitary quantum circuit that maps a standard state, assumed for
simplicity to be the computational basis state $|0\rangle$, to $|\Psi_{\rm
  prior}\rangle$. Deterministic updating is the topic of Section III.

\section{Probabilistic algorithms}
\label{sec:ProbabilisticAlgorithms}

As we have shown above there is in general no trace preserving
quantum operation that can transform all prior states
into the corresponding posterior state. To 
realize probabilistic Bayesian updating, we proceed as follows.
Define
\begin{equation} \label{definition:E_0}
E_1=C\sum_{h\in\set{S}_{\rm pr}}\sqrt{P(d|h)}\,|h\rangle\langle h|\,,
\end{equation}
where $C$ is a constant and $\set{S}_{\rm pr}$ is 
a set containing the support of the
prior probability distribution. We see that
\begin{equation}
E_1|\Psi_{\rm prior}\rangle\propto|\Psi_{\rm posterior}\rangle\,.
\end{equation}
For sufficiently small $|C|$, see Eq.~(\ref{BoundOnC}) below, 
one can view $E_1$ as an
element of a trace preserving quantum operation 
${\cal E}$ defined, for arbitrary $\rho$, by 
\begin{equation}
{\cal E}(\rho)=\sum_{k=0}^1E_k\rho E_k^\dag=\sum_{k=0}^1 p_k\rho(k)\,,
\end{equation} 
where
\begin{equation}
p_k=\Tr(E_k\rho E_k^\dag)\ \ \ \ {\rm and}\ \ \ \ 
\rho(k)=E_k\rho E_k^\dag/p_k\,.
\end{equation}
This decomposition shows that the operation ${\cal E}$
can be realized as a measurement with outcomes $k=0,1$, where
each outcome $k$ happens with probability $p_k$ and the
corresponding conditional density matrix is $\rho(k)$.
Substituting $\rho=|\Psi_{\rm prior}\rangle\langle\Psi_{\rm prior}|$
we see that the measurement outcome $1$ corresponds
to successful Bayesian updating. This
happens with probability
\begin{equation} \label{p0}
p_1=\langle\Psi_{\rm prior}|E_1^\dag E_1|\Psi_{\rm prior}\rangle
    =C^2\sum_{h}P(h)P(d|h)=C^2P(d)\,.
\end{equation}
In order to obtain a bound on $C$, we note that
\begin{equation} \label{E1_squared}
E_0^\dag E_0=\one-E_1^\dag E_1=\one-C^2\sum_{h\in\set{S}_{\rm pr}}P(d|h)\,|h\rangle\langle h|\,.
\end{equation}
Using the positivity of $E_0^\dag E_0$, we find
\begin{equation} 
C^2\leq\left(\sum_{h\in\set{S}_{\rm pr}} P(d|h)\,|\langle v|h\rangle|^2\right)^{-1}
\end{equation}
for any vector $|v\rangle$. 

Now let $h^*$ be such that $P(d|h^*)=\max_{h\in\set{S}_{\rm pr}} P(d|h)$. 
Since the above
condition is valid for any $|v\rangle$, one can choose 
$|v\rangle=|h^*\rangle$ and obtain
\begin{equation} \label{BoundOnC}
C^2\leq 1/\max_{h\in\set{S}_{\rm pr}}P(d|h)\,.
\end{equation}
Together with Eq.~(\ref{p0}) this gives an upper bound
on the success probability of Bayesian updating
\begin{equation} \label{SuccessProbabilityBound}
p_1\leq\frac{P(d)}{\max_{h\in\set{S}_{\rm pr}}P(d|h)}\,.
\end{equation}
In the next subsection we describe an explicit algorithm that achieves this bound. 

\subsection{Explicit algorithm}\label{subsec:ExampleAlgorithm}

The operation ${\cal E}$ can be realized as a modification of a procedure
proposed by Rudolph~\cite{RudolphPrivate} as follows. First we prepare the
product of the prior state and an auxiliary qubit state, $|\Psi_{\rm
  prior}\rangle|0\rangle$.  Then, using the classical algorithm for computing
$P(d|h)$, one can construct a quantum circuit $U_d$ that performs a
conditional rotation of an auxiliary qubit so that
\begin{equation} \label{U_d}
U_d |\Psi_{\rm prior}\rangle|0\rangle
=\sum_{h}\sqrt{P(h)}|h\rangle\Big(A_1(h)|0\rangle+B_1(h)|1\rangle\Big)\,,
\end{equation}
where 
\begin{equation} \label{A_1}
A_1(h)=c_1\sqrt{P(d|h)},\ \ \ B_1^2=1-A_1^2=1-c_1^2P(d|h)\,,
\end{equation}
and $c_1$ is a constant. Then measuring the auxiliary qubit
we obtain the desired state $|\Psi_{\rm posterior}\rangle|0\rangle$
with probability
\begin{equation}
p_1=c_1^2\sum_h P(h)P(d|h)=c_1^2P(d)\,.
\end{equation}
Looking at Eqs.(\ref{U_d}) and (\ref{A_1}) we can set
$c_1^2=1/\max_{h\in\set{S}_{\rm pr}}P(d|h)$. 
With this
setting, $p_1$ achieves the theoretical bound on the
success probability, Eq.(\ref{SuccessProbabilityBound}).

In the above algorithm, one can safely achieve the maximal
success probability only if the knowledge of
the value of $\max_{h\in\set{S}_{\rm pr}}P(d|h)$ 
is available. It is relevant to mention here
that the lack of such knowledge does not prevent
us from using the above algorithm, since we can always
use the trivial setting $c_1^2=1$. The price to pay is 
a smaller success probability.

An intermediate situation occurs if a nontrivial upper bound on
$P(d|h)$ is known, i.e., a constant 
$M$ such that $\max_{h\in\set{S}_{\rm pr}}P(d|h)<M<1$.
One can then set $c_1^2=1/M$, which improves the success probability compared
to the trivial setting.

\subsection{Iterative algorithm} \label{subsec:ExpensivePriors}

Let $M_1$ be an
upper bound on $\max_{h\in\set{S}_{\rm pr}}P(d|h)$.
Imagine that at the beginning we do not have
enough information about $P(d|h)$ and $P(h)$
to calculate a nontrivial value for $M_1$.
In other words, we have to assume that $M_1=1$.
Imagine also that we expect to acquire a 
better bound $M_2<M_1$ in the future. 
We will now address the following question: Can we run the
probabilistic algorithm of Sec.~\ref{subsec:ExampleAlgorithm}
first with the trivial bound $M_1=1$, and later with the improved bound
$M_2$, without reducing the overall success probability
that can be achieved by running the algorithm once with the bound $M_2$?
We will find that this is indeed the case.
This result remains true for a sequence of bounds, $M_k<M_{k-1}<\dots<M_1$.
Below we describe an iterative version of the above algorithm that
makes use of better bounds as they become available.

Consider the measurement part of the algorithm of
Sec.~\ref{subsec:ExampleAlgorithm}. 
If the measurement fails, which happens with probability  $1-p_1$, we
end up with the state
\begin{equation} \label{psi_1}
|\psi_1\rangle= \Big( N_1\sum_h\sqrt{P(h)}B_1(h)|h\rangle\Big) |1\rangle\,,
\ \ \ \ N_1^{-2}=1-c_1^2P(d)\,,
\end{equation}
where we might have set $c_1^2=1/M_1$ to maximize $p_1$.
Since we know the 
exact form of $|\psi_1\rangle$ we may attempt to achieve 
our original goal by
performing a transformation 
\begin{equation} \label{second_attempt}
|\psi_1\rangle\longrightarrow N_1\sum_h\sqrt{P(h)}B_1(h)|h\rangle
                              \Big(A_2(h)|0\rangle+\frac{B_2(h)}{B_1(h)}|1\rangle\Big)\,,
\end{equation}
where we set
\begin{equation}
A_2(h)=c_2\frac{\sqrt{P(d|h)}}{B_1(h)}\,,\ \ \ \ B_2^2=(1-A_2^2)B_1^2
                                                     =B_1^2-c_2^2P(d|h)\,,
\end{equation}
and $c_2$ is a constant. First of all, it is important to note
that this procedure should not be attempted
when $c_1^2$ was set to $1/M_1$, and $M_1$ is
still the best available bound. This is because
in the worst case there will be at least one hypotheses
$h^*$ which is present in the sum Eq.(\ref{second_attempt})
with $B_1(h^*)=0$ and $A_2(h^*)>1$. It follows that
the above procedure should only be applied if
a better bound $M_2>M_1$ became available (or when $c_1^2<1/M_1$). 
In this case, 
measurement of the auxiliary qubit
yields the desired state $|\Psi_{\rm posterior}\rangle|0\rangle$
with probability
\begin{equation}
p_2=N_1^2c_2^2\sum_hP(h)P(d|h)=\frac{c_2^2P(d)}{1-c_1^2P(d)}\,.
\end{equation}
Alternatively, with probability $1-p_2$, we may end up with the state
\begin{equation}
|\psi_2\rangle=\Big( N_2\sum_h\sqrt{P(h)}B_2(h)|h\rangle\Big) |1\rangle\,.
\end{equation}
This state is similar in structure to the state $|\psi_1\rangle$
so we may try to recover in the same way
by performing the transformation
\begin{equation}
|\psi_2\rangle\longrightarrow N_2\sum_h\sqrt{P(h)}B_2(h)|h\rangle
                              \Big(A_3(h)|0\rangle+\frac{B_3(h)}{B_2(h)}|1\rangle\Big)\,,
\end{equation}
followed by the measurements of the auxiliary qubit in complete analogy
with our earlier analysis. By continuing this procedure we obtain
the sequence of success probabilities $p_1,p_2,\dots$  
together with the coefficients $\{A_k^2\}$ and $\{B_k^2\}$. 
We have
\begin{equation} \label{AkBk}
A_k(h)=c_k\frac{\sqrt{P(d|h)}}{B_{k-1}(h)}\,,\ \ \ \ B_k^2=B_{k-1}^2-c_k^2P(d|h)\,,
\end{equation} 
and
\begin{equation} \label{p_k1}
p_k=\frac{c_k^2P(d)}{\langle B_{k-2}^2\rangle-c_{k-1}^2P(d)}\,,
\end{equation}
where $B_{-1}^2=B_0^2=1$, $c_0^2=0$ and
\begin{equation} \label{Baverage}
\langle B_k^2\rangle=\sum_h P(h) B_k^2(h)\,.
\end{equation}
The constants $\{c_k\}$ are the only free parameters in this
algorithm. As we have seen in the case $k=1$, 
the constants $\{c_k\}$ cannot be chosen
freely, and the optimal choice for them depends on the
sequence $\{M_k\}$.
From Eq.(\ref{AkBk}) we obtain
\begin{equation} \label{BfromCs}
B_k^2=1-P(d|h)\sum_{s=1}^k c_s^2\geq 0\,,
\end{equation}
and therefore
\begin{equation}
\sum_{s=1}^kc_s^2\leq\frac{1}{P(d|h)}\,.
\end{equation}
This condition must be satisfied for all $h$
in the support of the prior
and so we have
\begin{equation} \label{BoundOnCs}
\sum_{s=1}^kc_s^2\leq\frac{1}{\max_{h\in \set{S}_{\rm pr}} P(d|h)}\,.
\end{equation}
From Eqs.~(\ref{Baverage}) and (\ref{BfromCs})
we compute
\begin{equation}
\langle B_{k-2}^2\rangle=1-P(d)\sum_{s=1}^{k-2}c_s^2\,.
\end{equation}
Together with Eq.~(\ref{p_k1}), this implies
\begin{equation}
p_k=\frac{P(d)c_k^2}{1-P(d)\sum_{s=1}^{k-1}c_s^2}\,.
\end{equation}
The probability that the algorithm is not
successful after the $n$th stage is
given by
\begin{equation}
P_{\rm fail}^n=\prod_{k=1}^n(1-p_k)=1-P(d)\sum_{s=1}^nc_s^2\,,
\end{equation}
which gives the corresponding success probability
\begin{equation}
P_{\rm succ}^n=1-P_{\rm fail}^n=P(d)\sum_{s=1}^nc_s^2
\leq P(d)/\max_{h\in\set{S}_{\rm pr}}P(d|h)\,,
\end{equation}
where we used the inequality~(\ref{BoundOnCs}).
We see that the theoretical bound for
the overall success probability of
transforming one copy of the prior state
$|\Psi_{\rm prior}\rangle$ into one copy of the posterior
state
$|\Psi_{\rm posterior}\rangle$ is achieved for
as long as at some stage $n$ of the algorithm
we have
\begin{equation} \label{BoundOnCs2}
\sum_{s=1}^nc_s^2=\frac{1}{\max_{h\in \set{S}_{\rm pr}} P(d|h)}\,.
\end{equation}
Given the sequence of upper bounds $M_1>M_2>\dots>M_k$, and
assuming that the information in the first $k-1$ of them
was already used without success, the optimal value $c_k^2$ for the next 
iteration of the algorithm, which takes into account the bound $M_k$, can be
calculated as
\begin{equation}
c_k^2=\frac{1}{M_k}-\sum_{s=1}^{k-1}c_s^2=\frac{1}{M_k}-\frac{1}{M_{k-1}}\,.
\end{equation}

\section{Deterministic updating}
\label{sec:DeterministicAlgorithms}

In this section we will assume that the prior is given in the form of a
unitary quantum circuit, $U$, that maps the computational basis state
$|0\rangle$, to the prior state. Apart from the constraint
$U|0\rangle=|\Psi_{\rm prior}\rangle$, $U$ is arbitrary. We first give an
algorithm for the special case of hypothesis elimination and then show how to
extend it to two-valued and more general models.

\subsection{Hypothesis elimination}

Imagine the situation where each piece of data $d$ partitions the set of 
hypotheses $\set{H}$ into two subsets: $\set{H}_d$
containing all hypotheses that are consistent with $d$, and
$\set{H}\,\backslash\,\set{H}_d$ containing all hypotheses that are rejected
by the data $d$. This leads to a special case of Bayesian
updating
where $P(d|h)$ takes only two different values~\cite{Soklakov-0412},
\begin{equation}
P(d|h)=\left\{\begin{array}{ll}
               1/|\set{H}_d| & {\rm if\ }h\in\set{H}_d\,,\cr
                           0 & {\rm otherwise}\,,
                  \end{array}
                   \right.
\end{equation}
where $|\set{H}_d|$ is the number of hypotheses that are consistent with the
data $d$. The posterior state~(\ref{PosteriorState}) takes the simple form
\begin{equation}
|\Psi_{\rm posterior}\rangle
=N\sum_{h\in\set{H}_d}\sqrt{P(h)}|h\rangle\,,
\end{equation}
where $N$ is the normalization factor.

Using the given classical algorithm for computing $P(d|h)$, we define a
quantum oracle, $O_d$, as
\begin{equation}
O_d|h\rangle=\left\{\begin{array}{ll}
                     -|h\rangle & {\rm if\ }h\in\set{H}_d\,,\cr
                                         |h\rangle & {\rm otherwise}\,.
                  \end{array}
                   \right.
\end{equation}
Furthermore, let $\Pi$ be a conditional phase shift defined by
\begin{equation}
\Pi|h\rangle=\left\{\begin{array}{ll}
                     -|h\rangle & {\rm if\ }h\neq0\,,\cr
                      |h\rangle & {\rm if\ }h   =0\,.
                  \end{array}
                   \right.
\end{equation}
These operations are combined with $U$ to form an operation, ${\cal A}$,
  defined by \cite{Brassard}
\begin{equation}
{\cal A} = U^{-1} \Pi\,UO_d \;.
\end{equation}
The circuit for ${\cal A}$ is the basic block of the quantum algorithm to
prepare $|\Psi_{\rm posterior}\rangle$.

It will be convenient to rewrite the prior state~(\ref{PriorState})
in the form
\begin{equation} \label{Prior}
|\Psi_{\rm prior}\rangle=\sin\frac{\vartheta}{2}\;|\alpha\rangle
             +\cos\frac{\vartheta}{2}\;|\beta\rangle\,,
\end{equation}
where
\begin{equation}
|\alpha\rangle
= S_{\set{H}_d}^{-1/2}
   \sum_{h\in\set{H}_d} \sqrt{P(h)}\,|h\rangle\,,\ \ \ \ \ \
   S_{\set{H}_d}=\sum_{h\in\set{H}_d}P(h)\,, \label{SHd}
\end{equation}
\begin{equation}   
|\beta\rangle=S_{\set{H}\,\backslash\set{H}_d}^{-1/2}
   \sum_{h\in\set{H}\, \backslash \set{H}_d} \sqrt{P(h)}\,|h\rangle\,,\ \ \ \ \ \
   S_{\set{H}\,\backslash\set{H}_d}=\sum_{h\in\set{H}\,\backslash\set{H}_d}P(h)\,,
\end{equation}
and
\begin{equation} \label{SinVartheta}
\sin\frac{\vartheta}{2}=\sqrt{S_{\set{H}_d}}\;.
\end{equation}
The last equation shows that knowing the total
prior probability of the hypotheses that are
consistent with the data $d$ is equivalent
to knowing the value of $\vartheta$. 

It can now be shown that repeated application of the circuit 
${\cal A}$ takes $|\Psi_{\rm prior}\rangle$
through the sequence of states
\begin{equation}
{\cal A}^{k}|\Psi_{\rm prior}\rangle
= \sin\left(\frac{2k+1}{2}\vartheta\right)\,|\alpha\rangle
             +\cos\left(\frac{2k+1}{2}\vartheta\right)\,|\beta\rangle\,.
\end{equation} 
The number of times,
$T$, of applications of ${\cal A}$ that
achieve the required transformation, 
\begin{equation} \label{TBayesian}
{\cal A}^{T}|\Psi_{\rm prior}\rangle
=|\alpha\rangle=|\Psi_{\rm posterior}\rangle\,,
\end{equation}
is therefore
\begin{equation}
T=(\pi/\vartheta-1)/2\,.
\end{equation} 
If $T$ is not an integer, there are two possibilities. Either one uses the
closest integer approximation to $T$ and includes the effect of the noninteger
part in the fidelity analysis (see below), or one follows $\lfloor T\rfloor$
applications of ${\cal A}$ with one application of a modified version of
${\cal A}$ where phases are shifted by less than $e^{i\pi}$ in both $O_d$ and
$\Pi$ \cite{Tim}.

\begin{figure}[here]
\begin{center}
\epsfig{file=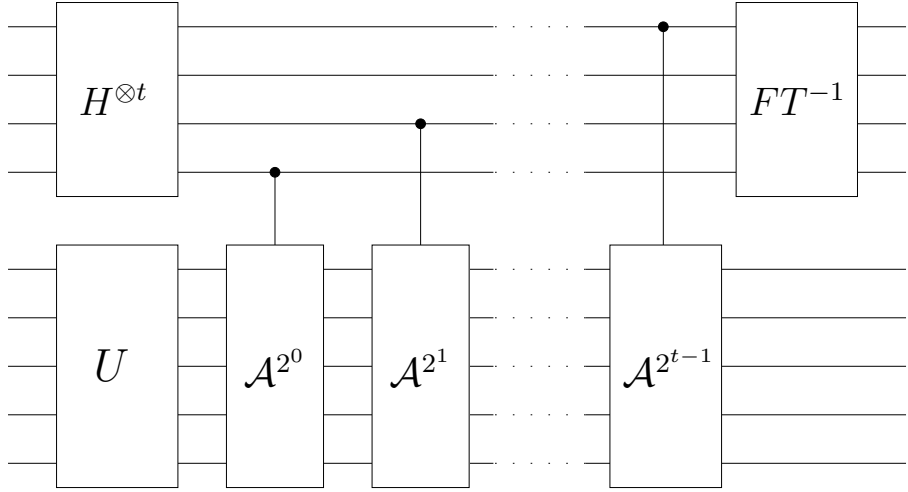,width=12cm}
\end{center}
\caption{This is the standard phase-estimation circuit applied to the
  hypothesis-elimination operator ${\cal A}$. A measurement of the upper
  $t$-qubit register returns the value of $\vartheta$ with an accuracy of $m$
  bits and a probability of success of at least $1-\epsilon$, where $m$ and
  $\epsilon$ are related to each other and to $t$ via the condition
  $t=m+\lceil \log(2+1/2\epsilon) \rceil$. The gates labeled $H^{\otimes t}$
  and $FT$ are the $t$-qubit Hadamard and quantum Fourier transforms,
  respectively.  }
\label{figure1}
\end{figure}

In order to compute the number of iterations, $T$, the value of $\vartheta$
must be known. To obtain $\vartheta$, a version of the standard phase
estimation algorithm \cite{Nielsen2000b} can be used as illustrated in
Figure \ref{figure1}.

To calculate the effect of an error in
the value of $\vartheta$  on the fidelity of the  Bayesian
transformation~(\ref{TBayesian}), we assume that there is an upper bound on
the absolute error, 
\begin{equation}
\Delta\vartheta\geq |\vartheta-\tilde{\vartheta}|\;,
\end{equation}
where $\tilde{\vartheta}$ denotes the approximate value. With the definition
$\tilde{T}=(\pi/\tilde{\vartheta}-1)/2$, the fidelity is
\begin{equation}
F=|\langle\Psi_{\rm posterior}|{\cal A}^{\tilde{T}}|\Psi_{\rm prior}\rangle|
=\sin\Big(\frac{2\tilde{T}+1}{2}\vartheta\Big)\,.
\end{equation}
Substituting $\vartheta=\tilde{\vartheta}\pm\Delta\vartheta$
and using the relation $(2\tilde{T}+1)\tilde{\vartheta}=\pi$
we obtain
\begin{equation} \label{FidelityBound}
F=\cos\Big(\frac{2\tilde{T}+1}{2}\Delta\vartheta\Big)
 =\cos\frac{\pi\Delta\vartheta}{2\tilde{\vartheta}}
 \geq 1-\Big(\frac{\pi\Delta\vartheta}{2\tilde{\vartheta}}\Big)^2\,.
\end{equation}

\subsection{Two-valued models} \label{sec:suppression}

A straightforward generalization of hypothesis elimination is provided by 
a two-valued conditional probability of the form
\begin{equation} \label{SingleStepModel}
P(d|h)=\left\{\begin{array}{ll}
               a_1 & {\rm if\ }h\in\set{H}_d\,,\cr
                           a_2 & {\rm otherwise}\,,
                  \end{array}
                   \right.
\end{equation}
where $a_1>a_2$ are constants, and $\set{H}_d$ is the set of
hypotheses favored by the data $d$. The {\em suppression coefficient\/}
$r=a_1/a_2$ measures how much hypotheses in $\set{H}_d$ are favored by the
data. As before, the prior state can be written in the
form, Eq.(\ref{Prior}),
\begin{equation}
|\Psi_{\rm prior}\rangle=\sin\frac{\vartheta}{2}\;|\alpha\rangle
             +\cos\frac{\vartheta}{2}\;|\beta\rangle\,,
\end{equation}
and for the posterior state we calculate
\begin{equation}
|\Psi_{\rm posterior}\rangle=\sqrt{a_1}\,\sin\frac{\vartheta}{2}\;|\alpha\rangle
             +\sqrt{a_2}\,\cos\frac{\vartheta}{2}\;|\beta\rangle\,.
\end{equation}
Normalization of the posterior state implies that
\begin{equation}
a_2=\frac{1}{r\sin^2(\vartheta/2)+\cos^2(\vartheta/2)}\,.
\end{equation}
Defining $\vartheta'$ so that
\begin{equation}
\cos\frac{\vartheta'}{2}= \sqrt{a_2}\,\cos\frac{\vartheta}{2}
=\frac{\cos(\vartheta/2)}{\sqrt{r\sin^2(\vartheta/2)+\cos^2(\vartheta/2)}}\,,
\end{equation}
the number of iterations $T$ necessary to transform $|\Psi_{\rm prior}\rangle$
into $|\Psi_{\rm posterior}\rangle={\cal A}^T|\Psi_{\rm prior}\rangle$ can then
be calculated as
\begin{equation}
T(\vartheta,r)=(\vartheta'/\vartheta-1)/2\,.
\end{equation}
It follows that knowledge of $\vartheta$ and the suppression coefficient $r$
is sufficient for a deterministic implementation of Bayesian updating with the
conditional distribution~(\ref{SingleStepModel}). As before, the
value of $\vartheta$ can be obtained using the algorithm of
figure~\ref{figure1}, and the same fidelity bound (\ref{FidelityBound}) can be
used.

\subsection{Bayesian updating: general models}

In this section we show how to generalize the above algorithm
to the case of Bayesian updating with a general model, 
i.e., a general conditional distribution $P(d|h)$.
The main idea  is to
represent $P(d|h)$ as a product of two-valued models
with known suppression coefficients. Bayesian updating
with $P(d|h)$ can then be viewed as a sequence of Bayesian
updatings for the two-valued models.

Let $C_k(h)$ be the coefficients in the binary expansion
of $\log_2P(d|h)$,
\begin{equation}
\log_2 P(d|h)=\sum_{k=1}^\infty C_k(h)\,2^{-k}\,.
\end{equation}
This allows us to express $P(d|h)$ as a product,
\begin{equation}
P(d|h)=\prod_{k=1}^{\infty}\sqrt[2^k]{2^{C_k(h)}}\,.
\end{equation}
Let $\set{H}_{d_k}$ be the set of hypotheses $\{h\}$ for which $C_k(h)=1$.
The $k$th term in this product is either $\sqrt[2^k]{2}$ or $1$ depending on
whether $h$ is in $\set{H}_{d_k}$ or not.  Bayesian updating with the
conditional probability $P(d|h)$ can therefore be viewed as a sequence of
stages corresponding to the acquisition of data from the sequence
$d_1,d_2,\dots$. At each stage, an updating step for a two-valued model as
described in the previous section is carried out. 

\section*{Acknowledgments}

We would like to thank Terry Rudolph for helpful
discussions. 
This work was supported in part by the European Union IST-FET project EDIQIP.

\end{document}